\documentclass[manuscript,screen]{acmart}
\AtBeginDocument{%
  \providecommand\BibTeX{{%
    \normalfont B\kern-0.5em{\scshape i\kern-0.25em b}\kern-0.8em\TeX}}}

\begin{document}

\title[Librarian-in-the-Loop]{Librarian-in-the-Loop: A Natural Language Processing Paradigm for Detecting Informal Mentions of Research Data in Academic Literature}

\author{Lizhou Fan}
\affiliation{%
  \institution{School of Information, University of Michigan}
  \city{Ann Arbor}
  \state{Michigan}
  \country{USA}
}

\author{Sara Lafia}
\affiliation{%
  \institution{ICPSR, University of Michigan}
  \city{Ann Arbor}
  \state{Michigan}
  \country{USA}
}

\author{David Bleckley}
\affiliation{%
  \institution{ICPSR, University of Michigan}
  \city{Ann Arbor}
  \state{Michigan}
  \country{USA}
}

\author{Elizabeth Moss}
\affiliation{%
  \institution{ICPSR, University of Michigan}
  \city{Ann Arbor}
  \state{Michigan}
  \country{USA}
}

\author{Andrea Thomer}
\affiliation{%
  \institution{School of Information, University of Michigan}
  \city{Ann Arbor}
  \state{Michigan}
  \country{USA}
}

\author{Libby Hemphill}
\affiliation{%
  \institution{School of Information \& ICPSR, University of Michigan}
  \city{Ann Arbor}
  \state{Michigan}
  \country{USA}
}

\renewcommand{\shortauthors}{Fan and Lafia, et al.}

\begin{abstract}
Data citations provide a foundation for studying research data impact. Collecting and managing data citations is a new frontier in archival science and scholarly communication. However, the discovery and curation of research data citations is labor intensive. Data citations that reference unique identifiers (i.e. DOIs) are readily findable; however, informal mentions made to research data are more challenging to infer. We propose a natural language processing (NLP) paradigm to support the human task of identifying informal mentions made to research datasets. The work of discovering informal data mentions is currently performed by librarians and their staff in the Inter-university Consortium for Political and Social Research (ICPSR), a large social science data archive that maintains a large bibliography of data-related literature. The NLP model is bootstrapped from data citations actively collected by librarians at ICPSR. The model combines pattern matching with multiple iterations of human annotations to learn additional rules for detecting informal data mentions. These examples are then used to train an NLP pipeline. The librarian-in-the-loop paradigm is centered in the data work performed by ICPSR librarians, supporting broader efforts to build a more comprehensive bibliography of data-related literature that reflects the scholarly communities of research data users.
\end{abstract}





\maketitle

\section{Introduction}
\label{sec:introduction}
Assessing the impact of research data requires knowledge of who has used data and for what purposes. Despite investments made to support research data preservation and curation, comparatively less is known about how data are reused. Recent work investigating the relationship between properties of datasets, curation decisions, and reuse has found that curated data are used more often \cite{hemphill2021properties} but there is more to learn about the context surrounding data reuse. 

The Inter-university Consortium for Political and Social Research (ICPSR) is a large social sciences data archive, which curates research data and maintains a collection of publications determined to have utilized data available at ICPSR, the ICPSR Bibliography of Data-related Literature.  It has strict inclusion criteria, only preserving references to resources that indicate substantial data reuse rather than passing mentions to datasets \cite{moss2015sharing}. ICPSR staff manually curate the Bibliography, which is time-consuming, given the volume of candidate citations returned for thousands of research studies in the archive.

Initiatives to measure data impact, such as Project COUNTER \cite{fenner2018code}, rely on formal data citation using persistent, unique identifiers (PIDs). The use of PIDs to reference datasets is an emerging practice, however \cite{mooney2011citing, park2018informal}. A study of the Dryad digital repository found that the share of articles referencing PIDs had grown from 69\% to 83\% between 2011 and 2014; however, the share of articles that included data identifiers in the works cited section remained low, under 10\% \cite{mayo2016location}. Many researchers reference data informally, for example by study name in sections of the main text such as methods, as well as in footnotes, tables, acknowledgements, and supplements \cite{park2018informal}.

Until a culture of formal data citation is established, studies of research data impact will rely on detection of informal data citations, which we describe here as ``mentions''. Machine learning (ML) and Natural Language Processing (NLP) have been used to support bibliometric research, for example, to match citation strings to complex research objects, like longitudinal studies and infer research fields and methods from citations \cite{mathiak2015challenges}.

Here, we describe the task of detecting informal data mentions and its application at ICPSR. We introduce the librarian-in-the-loop workflow for detecting informal references to research data in the full text of academic publications, which combines iterative steps in: 1) representing librarians' knowledge and experience as Named Entity Recognition (NER) annotations; and 2) building and training ML and NLP models. We conclude with a discussion of our work's implications for the development of impact metrics for research data.

\section{Librarian-in-the-Loop}
\label{sec:method}
Our librarian-in-the-loop workflow implements human-in-the-loop ML and NLP in the fields of digital library and archival science. The human-in-the-loop paradigm aims to optimize model training through the interaction between human knowledge and the entire machine learning process, from data preparation to model evaluation \cite{wu2021survey, monarch2021human}. 

Recent progress in conceptualizing and applying human-in-the-loop approaches provides implications and opportunities for solving the challenging NLP task, NER. Through interactive operations and potentially iterative evaluations, users can annotate samples, train the model, verify the accuracy, and update by annotating additional samples, which effectively reduces the task load while keeping up with or even exceeding the performance of benchmarks \cite{zhao2021human}. In few-shot learning settings for NER tasks, NER models trained on a small number of in-domain labeled data have proven to be a valid approach that can improve or outperform the baseline \cite{huang2020few}.

\begin{figure}
    \centering
    \includegraphics[width=.6\textwidth]{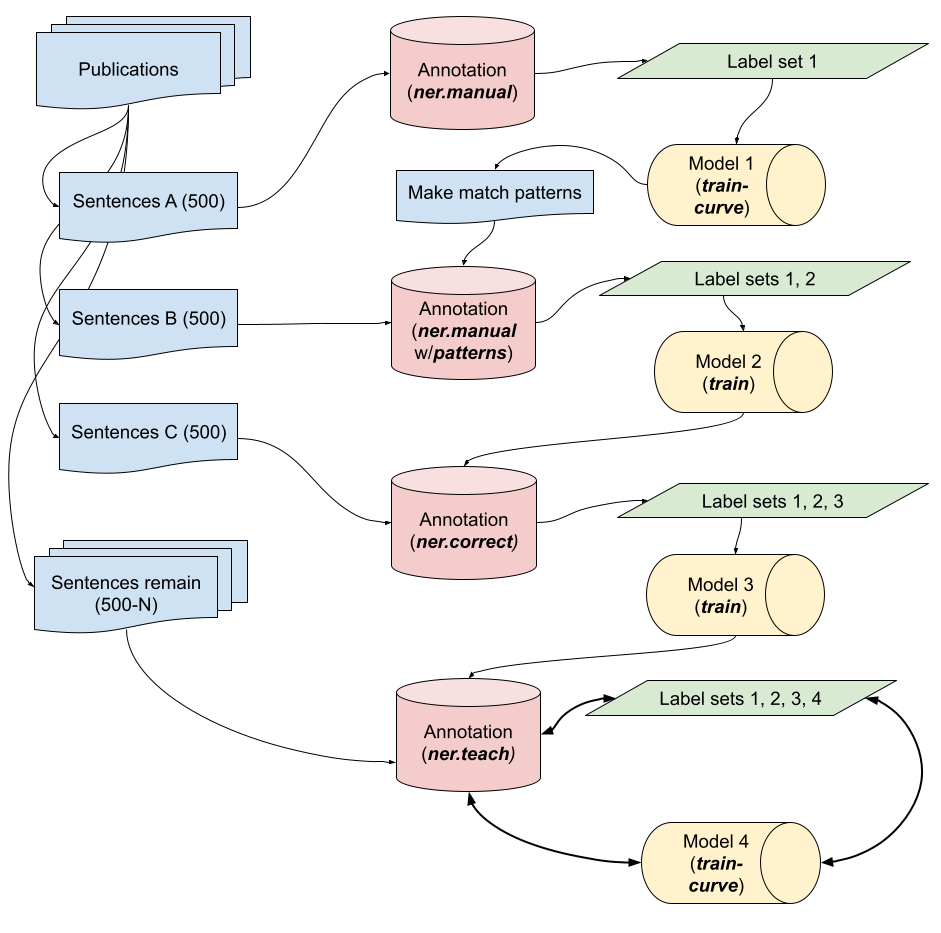}
    \caption{Librarian-in-the-loop: NER workflow for detecting data mentions}
    \label{fig:workflow}
\end{figure}

In particular, the librarian-in-the-loop workflow depends on: 1) librarians' digital curation expertise; and 2) computational models (ML and NLP). In our use case, we leverage this workflow to detect informal data mentions made to ICPSR studies in academic publications. As shown in Figure \ref{fig:workflow}, the librarian-in-the-loop workflow takes data citation candidate sentences (blue blocks) extracted from academic articles as the input. Through iterative annotation with different methods (red blocks), we create sets of NER labels (green blocks) of DATASET, which shows the text spans and the positional ranges of the mentioned datasets' names or acronyms. Using these cumulative labels (green blocks), we then train and improve the NER model for detecting data-citing spans in unseen sentences. 

\subsection{Data Prepossessing for Input Sentences}
\label{sec:preprocess}


We use two corpora of academic research articles to develop our workflow: 1) the ICPSR Bibliography of Data-Related Literature\footnote{\url{https://www.icpsr.umich.edu/web/pages/ICPSR/citations/}}; and 2) The Semantic Scholar Open Research Corpus (S2ORC)\footnote{\url{https://github.com/allenai/s2orc}}. The inputs (blue blocks in Figure \ref{fig:workflow}) for the workflow are sentences from the full-text of academic publications. Instead of having librarians manually rule out the irrelevant sentences to data citations, we use morphological patterns to select candidates and reduce the annotation load. Inspired by the rule-based method for detection of dataset names in scientific articles in \cite{heddes2021automatic}, we create a three-level candidate  extraction method for sentences defined with three possibility levels -- high (HIGH), middle (MID), and low (LOW) -- based on the observed patterns of datasets associated with a variety of fields and topics in the ICPSR Bibliography. The sentence extraction method can be described using keywords and regular expressions as:
\begin{itemize}
    \item HIGH -- keywords and their variations directly describing a DATASET, to be matched using regular expressions regardless of upper and lower cases, including  \verb/'(?:train|test|validation|testing|trainings?)\s*(?:set)'/, \verb/'data'/,  \verb/'data\s*(?:set|base)s?'/, \verb/'corp(us|ora)'/, \verb/'tree\s*bank'/, \verb/'collections?'/, \verb/'benchmarks?'/, \newline \verb/'surveys?'/, \verb/'samples?'/, \verb/'stud(y|ies)'/, \verb/'reports?'/, \verb/'census(es)?'/;
    \item MID -- the regular expression that can present acronyms of names of DATASET: \verb/\b[A-Z]{3,}s?\b/;
    \item LOW -- the regular expression that can present the full names of DATASET: \verb/([A-Z][a-z]+\s){2,}[A-Z][a-z]+/.
\end{itemize}
The detected sentences, with the span of the detected dataset citation spans, are provided together as input in the annotation steps. As an example, this bolded mention indicates a dataset: ``We also investigate individual-level black-white thermometer scores from waves of the \textbf{American National Election Survey (ANES)} from 1984 until 1998\dots".

\subsection{NER Annotation of DATASET}
For the iterative annotation process (red blocks in Figure \ref{fig:workflow}), we use Prodigy, an annotation tool for efficient machine teaching, and NER is one of its supporting features \cite{Prodigy:2018}. Through iteration, we use different built-in annotation methods to annotate NER in sentences by giving or changing the labels for spans, including:
\begin{itemize}
    \item \verb/ner.manual/ -- on a raw sentence, the annotator can highlight the entity span for the label (in our case, DATASET) if there is any detected entity;
    \item \verb/ner.correct/ -- on a sentence with pre-highlighted entity span(s), the annotator can correct the wrong entities by removing the highlight or add new span(s) by highlighting as in \verb/ner.manual/;
    \item \verb/ner.teach/ -- on a sentence with pre-highlighted entity span(s), the annotator can quickly conduct a binary annotation by clicking on ACCEPT the entity span(s) if every span is corrected detected or REJECT otherwise.
\end{itemize}
As Figure \ref{fig:workflow} shows, these three methods are used in the planned iterations in the librarian-in-the-loop workflow, where \verb/ner.manual/ is used twice. In the second iteration,  the \verb/ner.manual/ method is used together with \verb/patterns/ extracted from the first available model, which enhances the rule-based patterns in Section \ref{sec:preprocess}.

\section{Discussion}
\label{sec:discussion}

The librarian-in-the-loop workflow is designed as a digital curation aid for the multidisciplinary ICPSR team to expand the Bibliography. Through the detection of research data mentioned informally in academic literature, the team can computationally and efficiently identify candidate sentences in publications with high precision and low recall. Thus, staff's limited time can be used on later steps of curatorial actions to build a more comprehensive bibliography. 

This workflow, and its resulting NLP paradigm, can further support the analysis of data-related literature. Impact metrics based on a comprehensive bibliography better reflects the scholarly communities of data users through a more  complete representation of co-citation relations. The  analysis of data-related publications, including their fields of study and authorship information, can be used to develop impact metrics for research data. Data metrics have implications for researchers, who can gain new insights into the importance of data creation and reuse, as well as archivists, who seek to understand the impact of their curatorial decisions on data reuse.

From the perspective of ML and NLP, the curated model and its training labels are high quality benchmarks and datasets for related research in computational social science, especially in science of science studies. Building domain-specific corpora is typically difficult to organize and expensive to conduct. The librarian-in-the-loop workflow incorporates annotation into curation and results in new training data inputs for research fields related to scholarly communication. Using these data and models, in return, can support applications for data-related literature. For example, the annotated ICPSR Bibliography can support a data citation recommendation system  -- users can search for research ideas and receive recommendations based on published ICPSR studies and datasets.

We regard the librarian-in-the-loop workflow as a protocol that can broadly benefit researchers, librarians, and archivists who have limited funding, personnel, or computational resources. There is great potential to further apply this workflow in many data-related fields to encourage easier and more customizable data creation.

\bibliographystyle{ACM-Reference-Format}
\bibliography{Reference}

\end{document}